\documentclass[showpacs,preprintnumbers,amsmath,amssymb,amsfonts,latexsym,prl,twocolumn,
superscriptaddress]{revtex4}
\usepackage{graphicx}% Include figure files
\usepackage{dcolumn}% Align table columns on decimal point
\usepackage{bm}% bold math
\def\comment#1{}

\begin{document}
\title{Criticality in the $2+1$-dimensional compact
 Higgs model and fractionalized insulators}
\author{A. Sudb{\o}}
%\email{asle.sudbo@phys.ntnu.no}
\affiliation{Department of Physics, Norwegian University of
Science and Technology, N-7491 Trondheim, Norway}
\author{E. Sm{\o}rgrav}
%\email{eivind.smorgrav@phys.ntnu.no}
\affiliation{Department of Physics, Norwegian University of
Science and Technology, N-7491 Trondheim, Norway}
\author{J. Smiseth}
%\email{jo.smiseth@phys.ntnu.no}
\affiliation{Department of Physics, Norwegian University of
Science and Technology, N-7491 Trondheim, Norway}
\author{F. S. Nogueira}
%\email{nogueira@physik.fu-berlin.de}
\affiliation{Institut f\"ur Theoretische Physik,
Freie Universit\"at Berlin, Arnimallee 14, D-14195 Berlin, Germany}
\author{J. Hove}
%\email{joakim.hove@phys.ntnu.no}
\affiliation{Department of Physics, Norwegian University of
Science and Technology, N-7491 Trondheim, Norway}

\date{Received \today}

\begin{abstract}
We use a novel method of computing the third moment $M_3$ of the 
action of the $2+1$-dimensional compact Higgs model in the adjoint 
representation with $q=2$ to extract correlation length and specific 
heat exponents $\nu$ and $\alpha$ without invoking hyperscaling. 
Finite-size scaling analysis of $M_3$ yields the ratios 
$(1+\alpha)/\nu$ and $1/\nu$ separately. We find that $\alpha$ and 
$\nu$ vary along the critical line of the theory, which however exhibits 
a remarkable resilience of $Z_2$ criticality. We propose this novel 
universality class to be that of the quantum phase transition from a 
Mott-Hubbard insulator to a charge-fractionalized insulator in two 
spatial dimensions.
\end{abstract}


\maketitle
Modelling of strongly correlated systems plays a central role in trying to 
understand unconventional metallic states in cuprate perovskites and other 
systems, which do not conform to the Landau Fermi-liquid paradigm 
\cite{Review}. One avenue of research attempting to establish a theory 
of non-Fermi liquids in more than one spatial dimension, focuses 
attention on effective gauge theories of matter fields representing 
the charge of doped Mott-Hubbard insulators, coupled to compact gauge 
fields emerging from strong  constraints on the dynamics 
of the fermions \cite{IntroRef,Matsui,Laughlin,DHLee,Nagaosa}. Compact 
$U(1)$ gauge fields exhibit topological defects in the form of monopole 
configurations. It has been suggested that the unbinding of such monopoles 
may be relevant for spin-charge separation in strongly correlated systems 
\cite{Mudry,Matsui,Nagaosa} and for describing quantum antiferromagnets 
when fluctuations around the flux-phase are taken into account \cite{Affleck}. 
One often arrives at a description in terms of three-dimensional $d=3$ compact 
QED ($cQED_3$). A formulation of charge-fractionalization in terms of a $Z_2$ 
lattice gauge theory coupled to matter fields, has also been put forth 
\cite{RS,Wen,Senthil}. The above provides a link between important phenomena in 
condensed matter physics and deep issues in high-energy physics, such 
as confinement in QCD, with which $cQED_3$ shares two  essential features, 
namely confinement and chiral symmetry breaking. 

One lattice model arrived at in this context is the compact Higgs model defined 
by the partition function \cite{Savit,FradShe,Matsui,Nagaosa,Motrunich}
\begin{eqnarray}
Z & = & \int_{-\pi}^\pi\left[\prod_{j=1}^{N} \frac{d \theta_j}{2\pi}
\right]\int_{-\pi}^\pi\left[\prod_{j,\mu} \frac{d A_{j\mu}}{2\pi}\right]  
 \exp \left[ S \right] \nonumber \\ 
S  & = & \beta \sum_{j, \mu} [1-\cos(\Phi_{\mu j})] + \kappa  
\sum_{{\rm P}} [1-\cos({\cal A}_{\mu j})],  
\label{Model}
\end{eqnarray}
where $N$ is the number of lattice sites, $\sum_{{\rm P}}$ runs over the plaquettes 
of the lattice, $\Phi_{\mu j} \equiv \Delta_{\mu} \theta_j-q A_{j\mu}$, and 
${\cal A}_{\mu j} \equiv \varepsilon_{\mu\nu\lambda} \Delta_{\nu} A_{j\lambda}$. 
We use the variables $(x=1/[\kappa +1],y=1/[\beta+1])$ in discussing the 
possible phases of this model \cite{Savit}. In Eq. (\ref{Model}), $\theta$ 
is the phase of a scalar matter field with unit norm representing holons, 
$\Delta_{\mu}$ is a forward lattice difference operator in direction $\mu$, 
while $A_{j \mu}$ is a fluctuating gauge field enforcing the onsite 
constraints from strong correlations in the problem. 

When $q=0$, the matter field decouples from the gauge field. The model 
has  one critical point in the universality class of the $3DXY$ model, 
$y_c=0.688$ ($y_c=0.75$ in the Villain-approximation), while the pure 
gauge theory is permanently confined for all values of $\kappa$ 
\cite{Polyakov}. When $q=1$, Eq. (\ref{Model}) is trivial on the line 
$x=1,0<y<1$, with no phase transition for any value of $y$. On the line 
$0<x<1,y=1$ the matter field is absent and the theory is permanently 
confined \cite{Polyakov}. The phase-structure for $q=2,d=3$ was briefly 
discussed in Ref. \onlinecite{FradShe} and subsequently investigated 
numerically \cite{Bhanot}, the phase diagram is known, cf. Fig. 5 of 
Ref. \onlinecite{Bhanot}. When $y \to 0$ there is an Ising transition at 
$x_c=0.5678$, when $x \to 0$ there is a  $3DXY$ critical point at $y_c=0.688$. 
Moreover, a critical line $\beta_c(\kappa)$ connects these two critical limits. 
Above $\beta_c(\kappa)$ the system resides in a {\it deconfined-Higgs phase}, 
while below $\beta_c(\kappa)$ it resides in a {\it confined phase}. We also 
note that the model in Eq. (\ref{Model}) with $q=2$  recently was proposed 
as an effective theory of a microscopic boson lattice model exhibiting 
charge-fractionalized phases \cite{Motrunich}. The confined phase of Eq. 
(\ref{Model}) is interpreted as a Mott-Hubbard insulating phase, while 
the deconfined-Higgs phase is interpreted as a charge-fractionalized  
insulating phase \cite{Wen,Senthil,Motrunich}. 

No ordinary second order phase transition takes place in the case $d=3,q=1$ 
\cite{Osterwalder,FradShe}. However, two  of us have recently shown \cite{KNS}
that when 
matter is coupled to a compact gauge-field in a {\it continuum theory} and treating 
the topological defects of the gauge-field in an analogous manner to that done 
in Ref. \onlinecite{Polyakov}, the permanent confinement of the pure gauge 
theory is destroyed. A confinement-deconfinement transition may take place 
via a Kosterlitz-Thouless like unbinding of monopole configurations \cite{KNS} 
in {\it three dimensions} due to the appearance of an anomalous scaling dimension
of the gauge-field induced by critical matter-field fluctuations \cite{Herbut}. 
The role of an anomalous scaling dimension has also been studied recently 
at finite temperature, in pure compact QED in $d=3$ with no matter fields 
present \cite{Chernodub}. In both Refs. \onlinecite{KNS,Chernodub}, the 
appearance of an anomalous scaling dimension is crucial. The authors of Ref. 
\cite{Chernodub} recently also considered Eq. (\ref{Model}) with $q=1$ numerically,
\cite{Chernodub2}, finding a recombination of monopoles into dipoles 
connected by  matter strings, consistent with Ref.  \cite{KNS}. 
%However,
%no attempt was made to establish the KT-character of this transition.

Given the relevance of the case $q=2$ to current central issues in condensed 
matter physics \cite{RS,Wen,Senthil,Motrunich}, the universality class of the 
phase transition across the critical line for $q=2$ warrants attention. We 
therefore compute  the critical exponents $\alpha$ 
and $\nu$. Our results i) {\it demonstrate} that the critical behavior
found in the limits $\beta \to \infty$ ($Z_2$) and the limit
$\kappa \to \infty$ ($U(1)$) are not isolated points, and ii) 
{\it on balance suggest} that Eq. (\ref{Model}) is a {\it fixed-line} 
with non-universal $\alpha$ and $\nu$ depending on $(\beta,\kappa)$
rather than exhibiting a $Z_2$- and a $XY$ universality class
separated by a multicritical point. 

We  express  Eq. (\ref{Model}) as follows \cite{Savit,Nagaosa}
\begin{widetext}
\begin{equation}
Z  =  Z_0(\beta,\kappa) ~ \sum_{\{Q_j\}} \sum_{\{ J_{j\nu} \}} 
\delta_{\Delta_{\nu} J_{j\nu},q Q_j}
\exp \left[-4 \pi^2 \beta \sum_{j,k} ~ \biggl( J_{j\nu} 
~J_{k\nu} + \frac{q^2}{m^2} Q_jQ_k \biggr)D(j-k,m^2)\right],
\label{Loopgas0}
\end{equation}
\end{widetext}
where $\delta$ is the Kronecker-delta,
$D(j-k,m^2)=(-\Delta_{\lambda}^2 + m^2)^{-1} ~\delta_{jk}$,
and $m^2=q^2 \beta/\kappa$. $Z_0(\beta,\kappa)$ is the partition function for 
massive spin waves and will hereafter be omitted. Note the constraint 
$\Delta_{\nu} J_{j\nu}=q Q_j$ in the functional integral. Here $Q_j$ is 
the monopole charge on the dual lattice site number $j$, while $J_{j \nu}$ 
are topological currents representing segments of either open-ended strings 
terminating on monopoles, or closed loops \cite{Savit}.
 
In the limit $\beta \to \infty$ at fixed $\kappa$, Eq. (\ref{Loopgas0}) takes the form
\begin{equation}
Z=  \sum_{\{Q\}} \sum_{\{ J_{j\nu} \}}   
\delta_{\Delta_{\nu} J_{j\nu},q Q_j}\exp
\left(-\frac{2 \pi^2 \kappa}{q^2}  \sum_{j} ~  J^2_{j\nu}\right).
\label{Theory}
\end{equation}
This is the loop-gas representation of the global $Z_q$ spin model in the 
Villain approximation \cite{Savit1}. 
From Eq. (\ref{Theory}), it is seen that the cases $q=1$ and $q \neq 1$ are 
fundamentally different. For $q=1$, the summations over $\{Q_i\}$ may be 
performed to produce a unit factor at each dual lattice site, eliminating 
the constraint. Hence, $Z=\left(\vartheta_3(0,e^{-2 \pi^2 \kappa}) \right)^N$ 
where $ \vartheta_3$ is an elliptic Jacobi function. No
phase transition occurs at any value of $\kappa$ for $q=1$ when 
$\beta \to \infty$. For $q = 2$, a phase transition  survives \cite{FradShe}. 
In the language of Eq. (\ref{Theory}), this crucially depends on the presence 
of the constraint $\Delta_{\nu} J_{j\nu} = q Q_j$. For $q \neq 1$, summing over 
{\it all} values of $\{ Q_j \}$ still provides a remnant constraint ensuring 
a theory sustaining a phase transition.

\begin{figure}[htbp]
\centerline{\scalebox{0.50}{\rotatebox{0.0}{\includegraphics{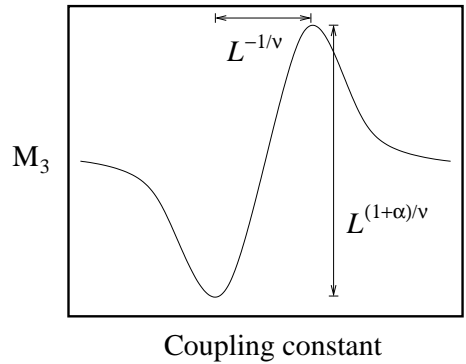}}}}
\caption{\label{Fig1} 
Generic third moment of action, $M_3$, showing how FSS is used to 
extract $\alpha$ and $\nu$.}
\end{figure}

We  compute the third moment $M_3$ of the action Eq. (\ref{Theory}),
$S = (2 \pi^2  \kappa/q^2)  \sum_{j} ~  J^2_{j\nu} \equiv 
(2 \pi^2 \kappa/q^2)~H$, with $M_n$ given by
\begin{equation}
M_n = \langle (H-\langle H \rangle)^n \rangle.
\label{Moment}
\end{equation}  
Using finite-size scaling (FSS) at the critical point, the peaks in $M_n$ scale with 
system size $L$ as $L^{(n-2+\alpha)/\nu}$. The width between the peaks in $M_3$
scales as $L^{-1/\nu}$, see Fig. \ref{Fig1}. Thus {\it both} $\alpha$ and $\nu$ 
are found from computing {\it one} quantity without using hyperscaling. Also, 
FSS of $M_3$ provides superior quality scaling compared to $M_2$ (specific heat), 
which unfortunately often is marred by significant confluent singularities. We have 
used multihistogram reweighting \cite{FerrSwend} of rawdata and jackknifing for error 
estimation in $M_3$ to perform FSS of the peak height and width between peaks. 

\begin{figure}[htbp]
\centerline{\scalebox{0.55}{\rotatebox{0.0}{\includegraphics{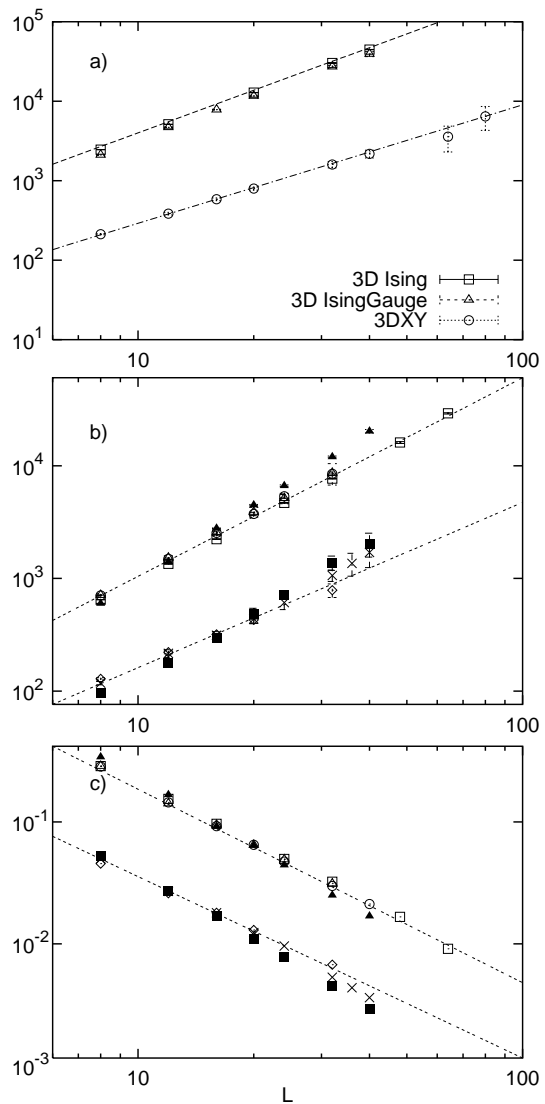}}}}
\caption{\label{Fig2} Log-log  scaling plots of peak-heights in $M_3$.
a) $3D$ Ising-, Ising gauge-, and $XY$-models. 
b) Eq. (\ref{Model}) with $q=2$,  
$\kappa/\beta=0.303 ~ (\bigcirc)$,
$\kappa/\beta=1.00 ~ (\bigtriangleup)$,
$\kappa/\beta=2.88 ~ (\blacktriangle)$,
$\kappa/\beta=3.47 ~ (\blacksquare)$,
$\kappa/\beta=4.05 ~ (\times)$,
$\kappa/\beta=5.14 ~ (\lozenge)$.
Also shown are results  for Eq. (\ref{Theory}) ($\square$). 
c) Scaling plots of width between peaks in $M_3$ for Eq. 
(\ref{Theory}), and Eq. (\ref{Model}) for $q=2$, legends as 
in b).}
\end{figure}

Before computing $M_3$ of Eq. (\ref{Theory}), and Eq. (\ref{Model}),
we perform benchmark Monte-Carlo simulations (MCS) on three well-known models.  
In Fig. \ref{Fig2} a) we show FSS results for the height of the peaks in $M_3$, 
defined analogously to Eq. (\ref{Moment}), for the $3D$ Ising-  and 
$3DXY$-models for $L=8,12,16,20,32,40,64$, with standard Metropolis updating. 
These are limiting cases of Eq. (\ref{Model}) (see below). Moreover (see 
below) the $3D$ Ising spin model is dual to the $3D$ Ising gauge theory
(IGT) \cite{Wegner}, and we have thus also computed $M_3$ for IGT.  
Any action must have $\alpha$ and $\nu$ identical with those of
its dual counterpart, since $\alpha$ and $\nu$ can be obtained directly
from scaling of the free energy, and are independent of the degrees
of freedom one chooses to describe the system in terms of. Our simulations 
bear this out with  precision, cf. Fig. \ref{Fig2} a), providing a 
 nontrivial quality check on them.

The system sizes we have used for the MCS on Eq. (\ref{Theory}) are $L^3$, 
with $L=8,12,16,24,32,48,64$, results are shown in Fig. \ref{Fig2} b)
($\square$-symbols). The allowed MC moves using Eq. (\ref{Theory}) are 
i) insertions of elementary loops made of vortex segments $J_{j \nu}= \pm 1$ and 
ii) insertions of open-ended vortex segments $J_{j \nu} = \pm q$  satisfying the 
constraint in Eq.  (\ref{Theory}). Up to $4 \cdot 10^6$ sweeps over the lattice 
have been used, with periodic boundary conditions in all directions.  

From Eq. \ref{Model}, the  limit $\beta \to \infty$, $\kappa$ fixed leads to 
the constraint $\Delta_{\mu} \theta_j - q A_{j\mu} = 2 \pi l_{j \mu}$ where 
$l_{j \mu}$ is integer valued. Substituting this into the gauge-field term 
in Eq. (\ref{Model}), we find   
\begin{equation}
Z = \prod_{j=1}^{N} \sum_{l_{j,\mu}=-\infty}^{\infty}
\exp \left[\kappa \sum_{{\rm P}} \left(1-
\cos (\frac{2 \pi {\cal L}_{\mu j}}{q}) \right) \right], 
\label{Model1}
\end{equation}
where ${\cal L}_{\mu j}=\varepsilon_{\mu\nu\lambda} \Delta_{\nu} 
l_{j \lambda} \in \mathbb{Z}$.  For $q=1$ the model is again seen to 
be trivial. Since Eqs. (\ref{Theory}) and (\ref{Model1}) are dual, and Eq. 
(\ref{Theory}) is a loop-gas representation of the global $Z_q$ theory 
while Eq. (\ref{Model1}) is the $Z_q$ lattice gauge theory, it follows
that the global and local $Z_q$ theories are dual in $d=3$ \cite{Guth}. 
Hence, the model Eq. (\ref{Model}) in the limit $\beta \to \infty$,  
$\kappa$ fixed, should have a ratio $(1+\alpha)/\nu$ consistent with 
the $Z_q$ spin model universality class, if the transition is 
continuous. 

For $q=2$, we have performed large-scale MCS and FSS analysis of $M_3$ 
of the action $S$ in Eq. \ref{Model}, written as 
$S= \beta ~ H_{\Phi} + \kappa ~ H_{{\cal A}}$, with 
$H_\lambda = \sum [1-\cos(\lambda_{\mu j})]$, cf. Eq. (\ref{Model}). A 
critical line $\beta_c(\kappa)$ separates a confined ($\beta < \beta_c$) 
and a Higgs-deconfined ($\beta > \beta_c$) state \cite{Bhanot}. We have 
used $L=8,12,16,20,24,32,40$, and up to $9 \cdot 10^6$ sweeps over the lattice 
with periodic boundary conditions in all directions. The critical line is 
crossed along the trajectory $\beta(\kappa)=\beta_c+a ~(\kappa-\kappa_c)$, 
where $(\beta_c,\kappa_c)$ is a point on the critical line. For the points 
at which $(1 + \alpha)/\nu$ has extrema, we use $a=(-1,1,\infty)$ to check 
that values for $\alpha$ and $\nu$ are not artifacts of how the critical 
line is crossed.

\begin{figure}[htbp]
\centerline{\scalebox{0.60}{\rotatebox{0.0}{\includegraphics{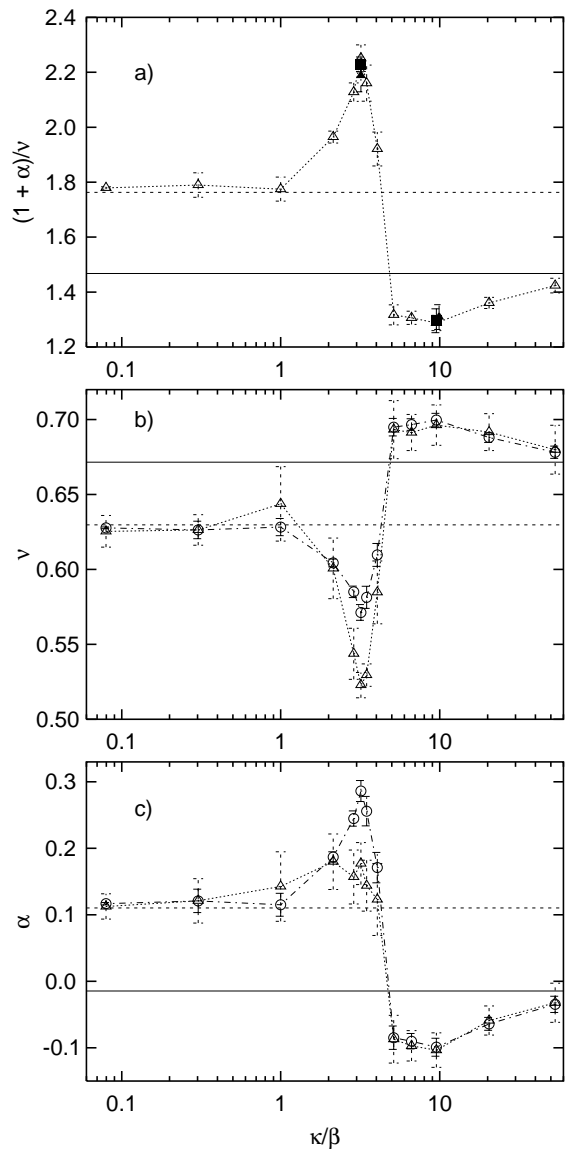}}}}
\caption{\label{Fig3} a)  $\Upsilon \equiv (1+\alpha)/\nu$  from 
FSS finite-size of $M_3$ for Eq. (\ref{Model}) for $q=2$. 
b) Same for the exponent $\nu$, computed directly from $M_3$ ($\triangle$)
and using results for $(1+\alpha)/\nu$ with hyperscaling 
($\bigcirc$). c) $\alpha$ as computed directly from $M_3$ 
($\triangle$) and using  hyperscaling ($\bigcirc$). The maximum and 
minimum in a) have been obtained by crossing the critical line along 
the trajectory $\beta(\kappa) = \beta_c + a ~ (\kappa - \kappa_c)$ 
with $a = \infty$ ($\triangle$), $a = 1$ 
($\blacksquare)$, and $a=-1$ ($\blacktriangle$) using $\beta_c = 
0.665, \kappa_c = 2.125$ (max.), and $\beta_c = 0.525, \kappa_c = 
5.0$ (min.). 
At the maximum,  
$\Upsilon(a=\infty) = 2.25 \pm 0.05$,
$\Upsilon(a=1)      = 2.23 \pm 0.03$,
$\Upsilon(a=-1)     = 2.19 \pm 0.06$. 
At the minimum, 
$\Upsilon(a=\infty) = 1.30 \pm 0.04$,
$\Upsilon(a=1)      = 1.31 \pm 0.05$,
$\Upsilon(a=-1)     = 1.29 \pm 0.03$. 
Dotted horizontal lines indicate $Z_2$-values, solid horizontal lines 
indicate $U(1)$-values.}
\end{figure}

In Fig. \ref{Fig2} b) we show scaling plots of the peaks in $M_3$ for 
Eq. (\ref{Model}) with $q=2$ for various values of 
$\kappa/\beta$ on the critical line, Fig. \ref{Fig2} c) shows 
corresponding scaling plots of the width between the peaks. 
From the finite-size scaling of the features in $M_3$, Fig. \ref{Fig1}, 
we extract the combination $(1+\alpha)/\nu$ as well as the exponent 
$1/\nu$ (and hence $\alpha$) along the  critical line, the results are 
shown in Fig. \ref{Fig3}. In Fig. \ref{Fig3} c), we also give values of 
$\alpha$ obtained directly from $M_3$ as well as using  $(1+\alpha)/\nu$ 
together with hyperscaling $\alpha = 2 - d \nu$. 
{\it We have checked that the extrema in $(1+\alpha)/\nu$ are not changed 
when the critical line is crossed in three very different directions, using 
$a=-1$, $a=1$ and $a = \infty$}. 

The results seem to rule out that  
$Z_2$- and $XY$-critical behaviors are isolated points at the extreme 
ends of the critical line. However, from Fig. \ref{Fig3}, it is feasible to 
suggest two types of universality, $Z_2$ and $XY$, separated at a multicritical 
point. We believe this to be ruled out by the strong deviation in $(1+\alpha)/\nu$
from $Z_2$- and $XY$-values at intermediate $\kappa/\beta$, which are
insensitive to $a$.  On balance, we thus conclude  that the model 
Eq. (\ref{Model}) defines a {\it fixed-line theory}, rather than exhibiting 
two scaling regimes separated by a multicritical point. However, the $Z_2$ 
character of the confinement-deconfinement transition persists  to 
surprisingly large values of $\kappa/\beta$ on the critical line, cf. 
Fig. 5 of Ref. \onlinecite{Bhanot}.  
Fixed-line theories in $2+1$ dimensions are known \cite{Note3}, 
and non-universal exponents imply the existence of marginal operators  
in Eq. (\ref{Model}), yet to be identified. 

Recently, Eq. (\ref{Model}) with $q=2$ was proposed as an effective 
theory for a microscopic model exhibiting a quantum phase transition 
from a Mott Hubbard insulator to a charge-fractionalized insulator in 
two spatial dimension \cite{Motrunich}. We thus propose that the zero 
temperature quantum phase transition from a Mott-Hubbard insulator to 
a charge-fractionalized insulator \cite{Senthil,Motrunich} is 
characterized by a {\it fixed-line theory} as given in Fig. \ref{Fig3},
but with remarkable $Z_2$ resilience.

A. S. and F.S.N. acknowledge support from the Norwegian Research Council
 and from the Humboldt Foundation. J. S and 
E. S. acknowledge support from the Norwegian University of Science and 
Technology (NTNU). A. S. thanks  H. Kleinert 
and the  FU Berlin for hospitality. We thank  NORDITA 
for hospitality, and  K. Rummukainen for providing 
the codes for multihistogram reweighting. Computations were
carried out at the Norwegian High Performance Computing Centre.

\end{document}